\documentclass[aps, prd, 12pt, showpacs, floatfix, twocolumn, nofootinbib, reprint, preprintnumbers, longbibliography]{revtex4-2}

\usepackage{multirow, amsmath, hyperref, graphicx, booktabs}
\usepackage{mathtools}
\usepackage{dcolumn}
\usepackage[dvipsnames]{xcolor}
\usepackage[final]{microtype}
\usepackage{booktabs}
\usepackage{array}

\usepackage{subcaption}
\hypersetup{
  colorlinks=true,
  citecolor=NavyBlue,
  linkcolor=red,
  urlcolor=NavyBlue,
  allbordercolors={0 0 0},
  pdfborderstyle={/S/U/W 1}
}

\usepackage{orcidlink}

\def \Hz	    {\mathrm{Hz}}
\def \msun	    {{M}_\odot}

\newcommand{\IITGn}{Department of Physics, Indian Institute of Technology Gandhinagar, Palaj Gandhinagar, Gujarat 382055, India.\vspace*{5pt}}

\begin{document}

\title{Probing dark matter halo profiles with multi-band observations of gravitational waves}

\author{\sc{Divya Tayelyani}\orcidlink{0000-0001-6146-7550}} 
\email{divya.tahelyani@iitgn.ac.in} \affiliation{\IITGn} 
\author{\sc{Arpan Bhattacharyya}\orcidlink{0000-0002-7933-6441}} 
\email{abhattacharyya@iitgn.ac.in} \affiliation{\IITGn}
\author{\sc{Anand S. Sengupta}\orcidlink{0000-0002-3212-0475}\vspace*{3pt}} 
\email{asengupta@iitgn.ac.in} \affiliation{\IITGn}

\begin{abstract} 
\textcolor{black}{In this paper, we evaluate the potential of multiband gravitational wave observations from a deci-Hz space-based detector and third-generation ground-based gravitational wave detectors to constrain the properties of dark matter spikes around intermediate-mass ratio inspirals. The presence of dark matter influences the orbital evolution of the secondary compact object through dynamic friction, which leads to a phase shift in the gravitational waveform compared to the vacuum case. Our analysis shows that the proposed Indian space-based detector GWSat, operating in the deciHz frequency band, provides the most stringent constraints on the dark matter spike parameters, as IMRIs spend a significant portion of their inspiral phase within its sensitivity range. While third-generation ground-based detectors such as the Einstein Telescope and Cosmic Explorer offer additional constraints, their contribution is somewhat limited, particularly for higher-mass systems where the signal duration in their frequency bands is shorter. However, for systems with detector-frame total masses $M_z < 400 \rm M_{\odot}$, Cosmic Explorer and Einstein Telescope could improve the estimation of the chirp mass, symmetric mass ratio, luminosity distance, and dark matter spike power-law index by more than $15\%$. Nonetheless, their impact on the constraint of spike density is minimal. These results highlight the crucial role of deciHz space-based detectors in probing dark matter interactions with gravitational wave sources.}

\end{abstract}
\maketitle 

\section{Introduction}
\label{sec:intro}
The direct observation of gravitational waves (GWs) has significantly enhanced our understanding of the universe, particularly in the study of black holes (BHs) and neutron stars (NSs). The detection of the first binary BH merger, GW150914, by the Laser Interferometer Gravitational-Wave Observatory (LIGO) in 2015 marked the beginning of GW astronomy~\cite{PhysRevLett.116.061102}, which has since led to a growing catalog of detected binary mergers.
Looking ahead, the next generation of GW detectors promises to further expand our capabilities. Ground-based GW observatories such as the Cosmic Explorer (CE)~\cite{Reitze:2019iox} and Einstein Telescope(ET)~\cite{Punturo:2010zz}, with their increased sensitivity, will enable the detection of fainter and more distant sources. Additionally, planned space-based detectors like LISA~\cite{LISA:2017pwj}, DECIGO~\cite{Yagi:2011wg}, and GWSat—a proposed Indian space-based GW detector—will aim to capture lower-frequency GW signals, particularly those emitted during the early inspiral phase of massive compact binary systems. These advancements in detector technology will provide a more complete picture of the universe's compact objects and allow us to probe new physics, such as the nature of dark matter (DM), by exploring the environments in which these binaries exist.
So far, most analyses of GW signals have assumed that source binaries evolve in vacuum environments, free from external perturbations. This assumption has served well for the initial detection of binary mergers by the LIGO-Virgo-KAGRA (LVK) collaboration~\cite{LIGOScientific:2018mvr, LIGOScientific:2020ibl, KAGRA:2021vkt, LIGOScientific:2018jsj, KAGRA:2021duu}, but it overlooks the possibility that many compact binaries could reside in dense astrophysical environments that might influence their orbital dynamics~\cite{Barausse:2014tra, Barausse:2014pra}. Recent theoretical developments suggest that compact binaries, particularly those involving massive black holes, may be embedded in DM halos or other dense astrophysical structures~\cite{Merritt:2002vj}. Such environments could measurably impact the emitted GWs, altering the observed waveforms and providing indirect evidence of the presence and distribution of dark matter~\cite{Macedo:2013qea, Eda:2013gg, Eda:2014kra, Yue:2017iwc,Yue:2018vtk,Hannuksela:2019vip,Cardoso:2019rou,Kavanagh:2020cfn,Coogan:2021uqv,PhysRevD.104.104042,PhysRevD.104.124082,Becker:2022wlo,Baryakhtar:2022hbu,Cardoso:2022whc,Singh:2022wvw,Destounis:2022obl,Kadota:2023wlm,Nichols:2023ufs, AbhishekChowdhuri:2023rfv, Bhattacharyya:2023kbh, Rahman:2023sof,Duque:2023seg,CanevaSantoro:2023aol,Speeney:2024mas,Aurrekoetxea:2024cqd,Zhang:2024ugv,Bertone:2024wbn,Spieksma:2024voy,Duque:2024mfw,Cheng:2024mgl}. 

Navarro, Frenk, and White (NFW) were the first to demonstrate through cosmological N-body numerical simulations that DM halos possess a universal equilibrium density profile, commonly known as the NFW profile~\cite{Navarro:1996gj}. Subsequently, Gondolo and Silk indicated that the adiabatic growth of supermassive black holes (SMBHs) could significantly alter the surrounding DM distribution, leading to the formation of DM overdensities known as DM spikes~\cite{Gondolo:1999ef}. However, host galaxy mergers and other astrophysical processes could disrupt these spike structures around SMBHs~\cite{Ullio:2001fb, Vasiliev:2008uz, Bertone:2005xv}. In contrast, intermediate-mass black holes (IMBHs), which have masses ranging from $10^2$ to $10^5 \rm M_{\odot}$, are less likely to experience such disruptions, making them ideal candidates for investigating DM spikes~\cite{Zhao:2005zr, Bertone:2005xz}.

The existence of DM spikes can influence the orbital evolution of compact binaries. When a stellar-mass compact object inspirals around an IMBH embedded in a DM spike, it experiences not only the gravitational pull of the BH but also a drag force known as dynamical friction, which results from interactions with the dark matter spike~\cite{Chandrasekhar:1943ys, Ostriker:1998fa, Kim:2007zb}. This dynamical friction alters the binary's inspiral rate, resulting in phase shifts in the gravitational waveform relative to what would be observed in a vacuum. Intermediate-mass ratio inspirals (IMRIs), which involve a stellar-mass object spiraling into an IMBH, are characterized by mass ratios in the range of $q = m_2/m_1 \sim 10^{-3}-10^{-5}$, with $m_1$ and $m_2$ representing the masses of the IMBH and the smaller companion, respectively. 

\textcolor{black}{Previous work by Kavanagh \textit{et al}.~\cite{Kavanagh:2020cfn} has demonstrated that these spikes are not static structures. Feedback mechanisms during binary inspiral dynamically reshape the DM profile. As a stellar-mass companion spirals into an IMBH, energy transfer via dynamical friction depletes the DM spike over time. Consequently, the DM density near the secondary object is reduced, which weakens dynamical friction and modifies the inspiral rate compared to the static case. When feedback on the DM distribution is taken into account, the dephasing caused by dynamical friction is significantly lower than in the static scenario, which is often discussed in the literature \cite{Karydas:2024fcn, Mukherjee:2023lzn, Cole:2022yzw, Kavanagh:2024lgq}. Nonetheless, the DM dephasing effect might still be substantial enough to be observed by future gravitational wave detectors~\cite{Cole:2022ucw, Coogan:2021uqv}. To assess the detectability of IMRIs in such evolving DM distributions, Ref.~\cite{Coogan:2021uqv} conducted a simulation-based analysis to evaluate their measurement prospects with LISA and their distinguishability from vacuum IMRIs. This study also proposed an analytical phenomenological model to describe the GW phase evolution in the frequency domain, facilitating Bayesian parameter estimation.}

Space-based detectors such as LISA, DECIGO, and GWSat are going to be particularly valuable for detecting the early inspiral phase of IMRIs, during which the influence of the DM spike is most significant. These detectors operate in a lower frequency range (millihertz to decihertz) compared to ground-based detectors, making them ideal for capturing the initial stages of the inspiral when the orbital velocity is lower and the DM environment has a significant impact on the binary's evolution. By combining data from both space-based and ground-based detectors—a technique known as multiband gravitational wave astronomy—it will be possible to track these systems across a broad range of frequencies, from the early inspiral phase to the final merger. This approach will significantly enhance parameter estimation and allow for more precise measurements of dark matter properties.

Multiband observations have been shown to dramatically enhance the precision of parameter estimation for compact binaries \cite{Sesana:2016ljz,Datta:2020vcj}. Previous studies have primarily focused on stellar-mass binary black holes like GW150914, showing that multiband observations can improve signal-to-noise ratios by several orders of magnitude compared to single-band observations~\cite{Sesana:2016ljz,Barausse:2016eii, Gnocchi:2019jzp, Grimm:2020ivq}. In the case of IMRIs, the advantages of multiband detection are even more pronounced. IMRIs may produce detectable signals in both space-based and ground-based detectors, with signal-to-noise ratios of the order of hundreds to thousands in the space-based band (e.g., GWSat) and tens in the ground-based band (e.g., ET/CE). This complementary capability allows for reduced uncertainties in the measurement of key parameters~\cite{Datta:2020vcj}, including the characteristics of DM spikes.

A recent study~\cite{CanevaSantoro:2023aol} conducted a Bayesian analysis of the first gravitational-wave catalog (GWTC-1) by LIGO-Virgo and found no significant evidence of environmental effects on the observed signals, but the results highlight the limitations of current detectors in detecting these subtle influences, particularly during the early inspiral phase where environmental effects are most likely to be identifiable. A numerical relativity study of GW150914-like events similarly suggests that environmental effects can mimic vacuum waveforms, potentially biasing parameter estimates \cite{Roy:2024rhe}.
Moreover, Ref.~\cite{Wilcox:2024sqs} demonstrated that using vacuum waveform templates for GW from inspirals in dark matter halos leads to biased parameter estimation, and suggests the necessity for an expanded parametrized post-Einsteinian (ppE) framework or more specialized DM waveform models.

\textcolor{black}{In this work, we investigate the potential of multiband gravitational wave observations to measure the properties of evolving dark matter spikes around IMRIs. We use the phenomenological model for GW phase for an IMRI inside a dynamic dark matter spike environment as described~\cite{Coogan:2021uqv}. We study how well the binary parameters, as well as dark matter parameters, could be constrained by combining the observations from an Indian deciHz space-based detector GWSat, and 3G ground-based detectors CE and ET using Fisher analysis framework. Our results show that GWSat plays a dominant role in constraining all parameters, particularly the dark matter spike properties, as the inspiral phase predominantly falls within its sensitivity band. The synergy between GWSat and 3G ground-based detectors strengthens error estimates, particularly for systems with $M_z < 400 \rm M_{\odot}$. In such cases, the addition of CE and ET improves constraints on chirp mass $\mathcal{M}_z$, symmetric mass ratio $\nu$, luminosity distance $D_L$, and dark matter spike index $\gamma_{\rm {sp}}$ by over $15\%$, depending on the source’s sky location. However, the ability to constrain $\rho_{\mathrm{sp}}$ remains largely driven by GWSat, highlighting its crucial role in uncovering dark matter properties through gravitational waves.} 

This paper is structured as follows: In Sec.~\ref{sec:DM_spike_model}, we describe the waveform models used for IMRIs in a DM environment and for parameter estimation. In Sec.~\ref{sec:Multiband_fisher}, we outline the multiband Fisher analysis methodology. Sec.~\ref{sec:results} presents the results obtained, which is followed by concluding remarks in Sec.~\ref{sec:conclusion}. Throughout the paper, we adopt the geometrized units, where G = c = 1.
\section{Gravitational waves from the evolution of the IMRI in dark matter spike}
\label{sec:DM_spike_model}

We consider an IMRI system comprising a stellar-mass compact object spiraling into an IMBH that is enveloped by a DM spike. The mass of the IMBH is denoted by $m_1$, while the mass of the small compact object is indicated by $m_2$. The total mass of the binary system is given by $M = m_1 + m_2$, and he orbital radius of the small compact object is symbolized by $r_2$. Initially, if the DM halo follows a cuspy profile characterized by $\rho(r)\propto r^{-\alpha_{\rm ini}}$ with $0 \leq \alpha_{\rm ini} \leq 2$, this distribution of dark matter will be altered due to adiabatic growth of the central IMBH, leading to the formation of a dense spike~\cite{Gondolo:1999ef}.

Following~\cite{Eda:2014kra}, we consider that the DM spike distribution is spherically symmetric DM with a simple power law:
\begin{equation}
   \rho_{\text{DM}}(r) = 
   \begin{cases} 
      \rho_{\text{sp}} \left( \frac{r_{\text{sp}}}{r} \right)^{\gamma_{\text{sp}}} & \text{for } r_{\text{in}} \leq r \leq r_{\text{sp}} \\ 
      0 & \text{for } r < r_{\text{in}} 
   \end{cases},
   \label{eq:static_density_distribution}
\end{equation}
where $r$ denotes the distance from the center of the IMBH. Following~\cite{Sadeghian:2013laa}, we define the inner radius of the spike as $r_{\text{in}} = 4 G m_1/c^2$. The maximum radius of the spike $r_{\mathrm{sp}}$ is determined by $r_{\mathrm{sp}}\sim 0.2 r_h$, where $r_h$ is the radius of the gravitational influence of the central BH defined by 
\begin{equation}
    \int_{r_{\text{in}}}^{r_{h}} \rho_{\text{DM}}(r) \, d^3r = 2m_1.
    \label{eq:r_h}
\end{equation}
From this, the maximum radius of the spike $r_{\text{sp}}$ is determined as
\begin{equation}
    r_{\text{sp}} \approx \left[ \frac{(3 - \gamma_{\text{sp}}) 0.2^{3 - \gamma_{\text{sp}}} m_1}{2 \pi \rho_{\text{sp}}} \right]^{1/3}.
    \label{eq:r_sp}
\end{equation}

The power-law index $\gamma_{\mathrm{sp}}$ that describes the final halo profile depends on the initial slope of the DM halo $\alpha_{\rm ini}$ which is related to the history of the formation of the IMBH via relation
\begin{equation}
    \gamma_{\rm sp} = \frac{9 - 2\alpha_{\rm ini}}{4 - \alpha_{\rm ini}}\,.
    \label{eq:gamma_sp}
\end{equation}
For the NFW profile, the initial slope is $\alpha_{\rm ini}=1$, henceforth the slope $\gamma_{\mathrm{sp}}=7/3=2.3\bar3$ is anticipated to form at the halo's center~\cite{Navarro:1996gj}. Following the setting used in~\cite{Eda:2014kra, Eda:2013gg}, we set $\rho_{\mathrm{sp}} = 226 \mathrm{M}_{\odot}/\mathrm{pc}^3$ and $\gamma_{\mathrm{sp}}=7/3$ throughout this paper. 

As a smaller compact object traverses a DM spike, the surrounding DM particles generate a disturbance that applies a drag force on it. This force reduces the smaller object's orbital speed, resulting in energy loss via dynamical friction (DF)~\cite{Chandrasekhar:1943ys}. This energy dissipation hastens the orbital decay of the smaller body, causing it to spiral inward at a rate faster than in the absence of matter. Apart from dynamical friction, the system also loses energy through gravitational wave emission. The overall rate at which the system's orbital energy decreases can be described as \cite{Coogan:2021uqv},
\begin{equation}
 \frac{\mathrm{d}E_{\text{orb}}}{\mathrm{d}t} = -\frac{\mathrm{d}E_{\text{GW}}}{\mathrm{d}t} - \frac{\mathrm{d}E_{\text{DF}}}{\mathrm{d}t}.
 \label{eq:energy_balance}
\end{equation}
The rate of energy loss due to GW emission for a circular orbit in the quadrupole approximation is expressed as \cite{Maggiore:2007ulw}:
\begin{equation}
    \frac{\mathrm{d}E_{\mathrm{GW}}}{\mathrm{d}t} = \frac{32G^4 M (m_1 m_2)^2}{5(cr_2)^5}.
    \label{eq:dE_GW}
\end{equation}
The energy loss because of dynamical friction is given by \cite{Coogan:2021uqv, Chandrasekhar:1943ys},
\begin{equation}
    \frac{\mathrm{d}E_{DF}}{\mathrm{d}t} = 4\pi (Gm_2)^2 \rho_{\mathrm{DM}}(r_2,\, t) \xi(v) v^{-1} \log \Lambda,
    \label{eq:dE_DF}
\end{equation}
where $v$ denotes the orbital velocity and the term $\xi(v)$ represents the fraction of dark matter particles with velocities lower than the orbital speed of the secondary object. As shown in~\cite{Nichols:2023ufs}, the factor $\xi$ is expressed as
\begin{equation}
    \xi = 1 - I_{1/2} \left( \gamma_{\rm sp} - \frac{1}{2}, \frac{3}{2} \right),
    \label{eq:xi}
\end{equation}
where $I_{1/2}$ is the regularized incomplete beta function which estimated to be $\xi \approx 0.58$ for $\gamma_{\mathrm{sp}} = 7/3$ as discussed in~\cite{Kavanagh:2020cfn}.
 The notation $\log \Lambda$ refers to the Coulomb logarithm, defined by $\Lambda = \sqrt{m_1/m_2}$ as in~\cite{Kavanagh:2024lgq, Kavanagh:2020cfn}. This logarithm accounts for the range of proximities from the smaller mass where gravitational interactions with DM particles become significant.

\textcolor{black}{Without accounting for feedback effects on the dark matter distribution, the density profile follows the static power-law form, $\rho_{\rm DM}(r_2, t) = \rho_{\rm DM}(r_2, 0)$, as given in Eq.~\ref{eq:static_density_distribution}. However, energy transfer from the secondary compact object to surrounding DM particles leads to transient depletion, causing the density profile to evolve over time. A comprehensive analysis requires joint evolution of the DM distribution and the dissipative dynamics of the binary, as detailed in~\cite{Coogan:2021uqv}. Additionally, Ref.~\cite{Coogan:2021uqv} developed a phenomenological waveform model for the time-domain phase of IMRIs in circular orbits at Newtonian order, incorporating only the dominant $l = 2,\, m = 2$ spherical harmonic mode. The parametrized form of this time-domain phenomenological phase model, calibrated to numerical simulations, expressed in terms of the GW frequency $f$ is given as~\cite{Coogan:2021uqv}:}
\textcolor{black}{
\begin{equation}
    \begin{aligned}
\Phi(f) \equiv &\,\Phi^{\rm V}(f) \\ 
& \times \left[1 - \eta y^{-\lambda} \left(1 - {}_2\rm F_1(1, \vartheta, 1+\vartheta, -y^{-5/(3\vartheta)}) \right) \right], \label{eq:parametrized_phase}
\end{aligned}
\end{equation}
where $y=f/f_t$ is a dimensionless frequency, $_2\rm F_1$ refers to the Gaussian hypergeometric function, and $\Phi^{\rm V}(f)$ defines the phase for the system in a vacuum with the chirp mass $\mathcal{M}=(m_1 m_2)^{3/5}/(m_1 + m_2)^{1/5}$ and the symmetric mass ratio $\nu = m_1 m_2/M^2$, as given by
\begin{equation}
\Phi^{\rm V} (f) = \frac{1}{16} \left( \frac{1}{\pi \mathcal{M} f} \right)^{5/3}\,.
\label{eq:phi_V}
\end{equation}
The parameters $(\eta, \lambda, \vartheta, f_t)$ are tuned for static and dynamic scenarios. For the parameters
\begin{equation}
\begin{aligned}
\eta & =1\,, \\
\lambda & =0\,, \\
\vartheta & =\frac{5}{11-2 \gamma_{\mathrm{sp}}}\,, \\
f_t & =f_{\mathrm{eq}}=c_f^{11-2 \gamma_{\mathrm{sp}}}\,,
\end{aligned}
\end{equation}
we obtain the Newtonian order GW phase $\Phi^{\rm S}(f)$ for an IMRI in a static DM spike. Here, $f_{\rm eq}$ is the frequency at which the energy loss rates from GWs and dynamical friction become equal. The coefficient $c_f$ is defined by \cite{Coogan:2021uqv},
\begin{equation}
    c_f = \frac{5}{8m_1^2} \pi^{\frac{2(\gamma_{\mathrm{sp}} - 4)}{3}}  (m_1 + m_2)^{\frac{1 - \gamma_{\mathrm{sp}}}{3}} r_{\mathrm{sp}}^{\gamma_{\mathrm{sp}}} \xi \rho_{\rm sp} \log \Lambda .
    \label{eq:cf}
\end{equation}
By choosing the parameters to be 
\begin{equation}
\begin{aligned}
\eta & =\frac{5}{8-\gamma_{\mathrm{sp}}}\left(\frac{f_{\mathrm{eq}}}{f_b}\right)^{\left(11-2 \gamma_{\mathrm{sp}}\right) / 3}\,, \\
\lambda & =\frac{6-2 \gamma_{\mathrm{sp}}}{3}\,, \\
\vartheta & =1\,, \\
f_t & =f_b\,,
\end{aligned}
\end{equation}
one obtains the GW phase $\Phi^{\rm D}(f)$ for the IMRI in a dynamic DM case at the Newtonian order. The scaling relation, as well as the empirical formula for the break frequency $f_b$ are obtained in~\cite{Coogan:2021uqv}.}

\begin{figure}
    \centering
    \includegraphics[width=0.85\linewidth]{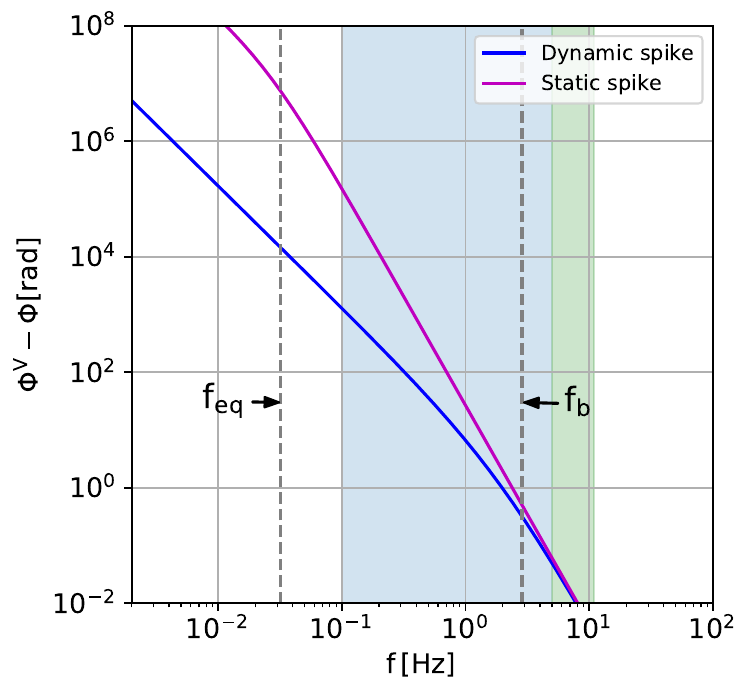}
    \caption{Following \cite{Coogan:2021uqv}, we have shown the dephasing $\Delta\Phi$ vs frequency for IMRIs in static and dynamic DM spike. The system considered here  $(m_1, m_2, \rho_{\mathrm{sp}}, \gamma_{\mathrm{sp}}) = (400 \mathrm{M}_{\odot}, 1.4 \mathrm{M}_{\odot}, 226 \mathrm{M}_{\odot}/\mathrm{pc}^3, 2.\bar{3})$ is different from \cite{Coogan:2021uqv}. The blue and magenta dashed lines denote the break frequencies for dynamic and static DM spikes, respectively. The blue and green shaded area shows the frequency bandwidth for which the binary appears in the space-based detector GWSat and the ground-based detector CE. The dephasing is shown up to the frequency at ISCO.} 
    \label{fig:dephasing}
\end{figure}

\textcolor{black}{Due to the dynamic friction effect, the inspiral process will involve fewer cycles before coalescence, resulting in a gradual phase difference compared to a vacuum scenario. This phenomenon is referred to as dephasing, which can be mathematically expressed as $\Delta \Phi = \Phi^{\rm V} - \Phi^{\rm}$. Fig.~\ref{fig:dephasing} depicts the GW phase difference, $\Delta \Phi$, between an IMRI in a vacuum and one within a DM spike. The dephasing $\Delta\Phi(f)$ nearly follows a broken power law for both static and dynamic DM spike scenarios, characterized by a distinct break frequency in each case \cite{Coogan:2021uqv}. The blue and green shaded regions in Fig.~\ref{fig:dephasing} represent the frequency ranges where the binary system is detectable by the space-based GWSat detector $([0.1, 5] \rm Hz)$ and the ground-based Cosmic Explorer $([5, f_{\rm ISCO}]\, \rm Hz)$, respectively. It is noted that $\Delta \Phi$ increases at low frequencies, which corresponds to larger binary separation, making this effect most prominent in the GWSat band. Therefore, it is crucial to consider observations from space-based detectors, which are sensitive in the mili-Hz and deci-Hz frequency ranges, in precisely constraining DM spike parameters such as $\rho_{\rm sp}$ and $\gamma_{\rm sp}$.}

\textcolor{black}{For frequencies  $f \gg f_b$, the system’s behavior closely resembles that of the same binary in a static DM spike. However, for  $f \ll f_b$, the dark matter halo can be significantly depleted at the orbital radius, reducing the effect of dynamic friction force relative to the static dark matter spike and, therefore, slowing down the inspiral process. This might greatly influence the effect of DM on the emitted GWs. Consequently, for binaries with a mass ratio $q > 10^{-3}$, it becomes essential to account for the feedback effect on the dark matter distribution. Since all binaries considered in this work have $q > 10^{-3}$, we adopt the dynamic spike profile in our analysis.}

\subsection{Gravitational-wave strain in the frequency domain}
The binary system generates GWs due to variations in its quadrupole moment. Using the stationary phase approximation (SPA),  the Fourier transform of the plus and cross-polarization waveforms can be expressed as~\cite{Maggiore:2007ulw}
\begin{equation}
    \tilde{h}_{+, \times} (f) = A_{+, \times}(f)\, e^{i\,\Psi(f)}
    \label{eq:GW_strain}
\end{equation}
where the amplitudes of two polarizations at leading order are given by
\begin{align}
    A_{+}(f) =& \frac{1}{D_L}\frac{(1+ \cos^2{\iota})}{2} h_0(f),\\
    A_{\times}(f) =& \frac{1}{D_L}\cos{\iota} \, h_0(f),\\
    h_0(f) =& \sqrt{\frac{5}{24}} \frac{\mathcal{M}^{\frac{5}{6}}f^{-\frac{7}{6}}}{\pi^{\frac{2}{3}}}   .
\end{align}
In these expressions, $D_L$ is the luminosity distance to the source, and $\iota$ represents the binary's inclination angle. The frequency domain SPA phase $\Psi(f)$ is related to the time domain phase in Eq.~\ref{eq:parametrized_phase} ($\Phi(t(f)$), as a function of frequency by~\cite{Coogan:2021uqv},
\begin{equation}
    \Psi(f) = 2\pi f (t_c + t(f)) - \phi_c -\frac{\pi}{4} + \Phi(f). 
    \label{eq:Psi_D}
\end{equation}
where $t_c$ is the coalescence time, and $\phi_c$ is the phase at coalescence.  The time evolution from an arbitrary initial frequency is given by 
\begin{equation}
    t(f) = \frac{1}{2\pi} \int \frac{\mathrm{d}f'}{f'} \frac{\mathrm{d} \Phi}{\mathrm{d} f}.\label{eq:t_of_f}
\end{equation}
\textcolor{black}{The frequency-domain phase for the dynamic dark matter case, derived from the time-domain phase expression in Eq.~\ref{eq:parametrized_phase}, is given by~\cite{Coogan:2021uqv, Wilcox:2024sqs}
\begin{equation}
\begin{aligned}
\Psi= & \Psi^{\rm V}(f)-\frac{5 \eta}{16\left(3 \bar{\lambda}^2+\bar{\lambda}-2\right)}\left(\pi \mathcal{M} f_t\right)^{-5 / 3} y^{-\bar{\lambda}} \\
& \times\left[{ }_2 \rm F_1\left(1, \frac{3(\bar{\lambda}+1)}{5} ; \frac{3 \bar{\lambda}+8}{5} ;-y^{-5 / 3}\right)\right. \\
& \left.-\frac{3}{5}(\bar{\lambda}+1) y^{5 / 3} \log \left(y^{-5 / 3}+1\right)+\frac{3}{5} \bar{\lambda}-\frac{2}{5}\right] .
\end{aligned}
\end{equation}
$\Psi^{\rm V}(f)$ represents the phase for the system in vacuum, given to leading post-Newtonian (PN) order by, 
\begin{equation}
    \Psi^{\rm V}(f)=2 \pi f t_c-\phi_c-\frac{\pi}{4}+\frac{3}{128}(\pi \mathcal{M} f)^{-5 / 3}
\end{equation}
and $\bar{\lambda}$ is defined as
\begin{equation}
  \bar{\lambda}=\lambda+\frac{5}{5}=\frac{11-2 \gamma_{\mathrm{sp}}}{3} \,. 
\end{equation}
}

The strain of GWs from the IMRI system measured by the detector in the frequency domain is given by the linear combination of the two GW polarizations $h_{+}$ and $h_{\times}$~\cite{Maggiore:2007ulw},
\begin{equation}
    \tilde{h}(f) = e^{i 2 \pi f \Delta t} \left ( \, F_{+}(\alpha, \delta, \psi) \,\tilde{h}_{+}(f) + F_{\times}(\alpha, \delta, \psi) \, \tilde{h}_{\times}(f) \right ).
    \label{detector_response}
\end{equation}
Here, $\Delta t$  is the time delay between the arrival of the GW signal at the Earth’s center and its arrival at the detector location. The antenna response functions $F_{+}$ and $F_{\times}$ determine the sensitivity of the detector to plus and cross polarizations, respectively. These functions are dependent on the sky position of the source, right ascension $\alpha$ and declination $\delta$, as well as polarization angle $\psi$.

\textcolor{black}{When estimating the parameter uncertainties using Fisher analysis, we account for the impact of Earth's rotation on the detector's antenna patterns and response. This consideration is particularly important for GW signals that persist for a long time within the detector's bandwidth, as the source's position in the sky can change significantly because of Earth's rotation relative to the detector. The antenna pattern functions change periodically with a period equal to one sidereal day. To incorporate this, we use the time-frequency correspondence relation Eq.~\ref{eq:t_of_f} within SPA and compute the time-dependent quantities in the response function. Using Eq.~\ref{eq:t_of_f}, the time corresponding to any instantaneous frequency $f$ before the merger can be computed. Further details on these calculations can be found in ~\cite{Zhao:2017cbb}. Ref.~\cite{Nair:2018bxj} also outlines the formalism needed to include the time dependence of antenna pattern functions resulting from Earth's rotation.}
\section{Multiband parameter estimation}
\label{sec:Multiband_fisher}
\subsection{Multiband Fisher analysis}
The Fisher matrix is a widely-used approach for predicting the statistical uncertainties in GW parameter estimation when the models for both signal and noise are known~\cite{Cramer1946, Cutler:1994ys,Vallisneri:2007ev}. In our analysis, we utilize the Fisher matrix approach to determine the $1\sigma$ uncertainties on both vacuum GR and dark matter model parameters. 

The GW strain data $s(t)$ recorded by a detector is typically modeled as a combination of the true gravitational wave signal $h(t, \boldsymbol{\theta})$ embedded in additive detector noise $n(t)$:
\begin{equation}
    s(t) = h(t, \boldsymbol{\theta}) + n(t).
\end{equation}
The likelihood function for the observed detector data, assuming stationary zero-mean Gaussian noise, is given by
\begin{equation}
\mathcal{L} \left(\boldsymbol{\theta}\right) =
\exp \left(-\frac{1}{2} 
\left \langle s-h\left(\boldsymbol{\theta}\right) 
    \mid s-h\left(\boldsymbol{\theta}\right)
    \right\rangle 
    \right) .
\end{equation}

The Fisher Information Matrix (FIM) is defined as
\begin{equation}
    \Gamma_{pq} = -\mathrm{E}\left[\frac{\partial^2 \ln\mathcal{L}(\boldsymbol{\theta})}{\partial \theta_p \partial\theta_q}\right],
    \label{eq:FIM}
\end{equation}
where E denotes the expectation value over the 
likelihood distribution $\mathcal{L} \left(\boldsymbol{\theta}\right)$. 
The Fisher information quantifies the amount of information a random variable (in our case, the noisy GW signal as observed in the detectors) provides about an unknown parameter or set of parameters. It may be seen as the curvature of the log-likelihood curve, with a high value indicating sharply peaked likelihood distribution, implying less variance on the maximum likelihood estimate of the parameters; and vice versa. One appeals to the Cram\'er–Rao~\cite{Rao1945} bound according to which, the inverse of the Fisher information provides a lower bound on the variance of any unbiased estimator of the parameters.
%
As shown in~\cite{HELSTROM1968249}, for a signal buried in stationary Gaussian noise, the FIM above given by Eq.~\eqref{eq:FIM} can be computed using the partial derivatives of the frequency-domain GW waveform with respect to the parameters:
\begin{equation}
    \Gamma_{pq} = 
    \left\langle \frac{\partial\tilde{h}(f)}{\partial \theta_{p}},\frac{\partial\tilde{h}(f)}{\partial \theta_{q}} \right\rangle
    \label{eq:Fisher_matrix} ,
\end{equation}
where $\tilde{h}(f, \boldsymbol{\theta})$ is the frequency-domain GW waveform template defined by the set of parameters $\boldsymbol{\theta}$. The inner product between two functions, $a(f)$ and $b(f)$, weighted by the detector noise, is defined as:
\begin{equation}
    \langle a,b\rangle=4 \mathcal{R}\int_{f_{\rm low}}^{\rm f_{\rm high}}\frac{a(f)\,b^*(f)+a^*(f)\, b(f)}{S_n(f)}\,df \,,
\end{equation}
where and ``$*$'' represents the complex conjugation, and $S_n(f )$ is the one-sided power spectral density (PSD) of the noise. 
The signal-to-noise ratio (SNR) for a given $\tilde{h}$ and a specific detector PSD $S_n(f)$ is given by
\begin{equation}
    \rho = \langle \tilde{h}, \tilde{h} \rangle^{1/2},
\end{equation}
which indicates the relative strength of the detector response in relation to the detector noise. In the limit of high SNR, the inverse of the FIM provides the error covariance matrix $\Sigma_{pq}=(\Gamma_{pq})^{-1}$ and the square root of the diagonal elements of $\Sigma_{pq}$ gives the 1$\sigma$ error for each parameter: $\sigma_p = \sqrt{\Sigma_{pp}}$. In the case of multiband observations, where signals are detected jointly by both a space-based detector and a ground-based detector, the total FIM is obtained by summing the individual FIM from each detector \cite{Datta:2020vcj}:
\begin{equation}
    \Gamma^{\mathrm{Combined}}_{pq} = \sum_{j} \Gamma_{pq}^{j},
\end{equation}
where $\Gamma_{pq}^j$ indicates the Fisher matrix associated with the j-th detector. The SNR for the network of detectors is defined as
\begin{equation}
    \rho^{\rm Combined}=\sqrt{\sum_j {\rho^j}^2},
\end{equation}
where $\rho^j$ denotes the SNR of j-th detector.
Because these detectors are independently measuring the same event but under different conditions, the signals they capture are essentially independent pieces of data. This independence allows us to sum up their contributions. 
The covariance matrix for the combined measurement and the corresponding error estimate is given as  \cite{Datta:2020vcj},
\begin{align}
    \Sigma^{\mathrm{Combined}}_{pq} =& (\Gamma^{\mathrm{Combined}}_{pq})^{-1}, \\
    \sigma^{\mathrm{Combined}}_{p} =& \sqrt{\Sigma^{\mathrm{Combined}}_{pp}} .
\end{align}
When we sum the Fisher matrices from multiple detectors, the total information about the parameters increases, and as a result, the combined covariance matrix $\Sigma^{\mathrm{Combined}}_{pq} = (\Gamma^{\mathrm{Combined}}_{pq})^{-1}$ will typically have smaller diagonal elements than the covariance matrix of any individual detector. Hence, the uncertainties on the estimated parameters are reduced.
\begin{figure*}[!ht]
    \centering
    \includegraphics[width=0.75\textwidth]{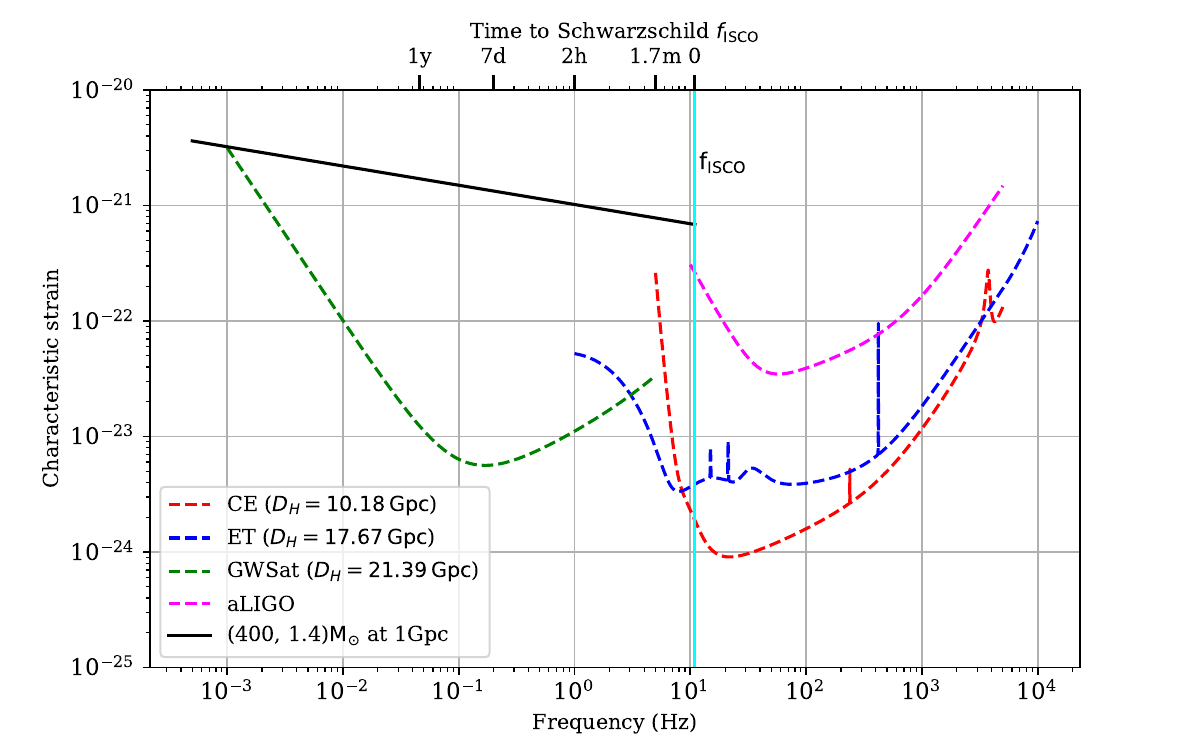}
    \caption{The dashed lines represent the characteristic strain of noise for GWSat, aLIGO, CE, and ET. The solid black line indicate the characteristic strain of signal $(2 f | \Tilde{h}( f )|)$ of an IMRI system in DM spike, with detector-frame masses 400$\mathrm{M}_{\odot}$ and 1.4$\mathrm{M}_{\odot}$ at 1 Gpc. $f_{\mathrm{ISCO}}$ for the considered system is shown by the cyan solid line. $D_H$  represents the horizon distance, which is the maximum observable distance for this IMRI system oriented face-on with respect to the detectors, at an SNR threshold of 10 for each detector. We have reproduced these curves using the PSDs for various detectors from \cite{PrivateCommMitra2024,Srivastava:2022slt, Hild:2010id, 
 aLIGO_design_sensitivity}. }
    \label{fig:noise_sensitivities}
\end{figure*}

\subsection{Detector configurations}

We now focus on evaluating the detectability of GW signals originating from compact binary systems situated within dark matter environments, utilizing a consortium of proposed space-based detector GWSat, and 3G ground-based detectors, which include Cosmic Explorer (CE) and the Einstein Telescope (ET). This network of detectors is crucial for multiband GW observations, as it facilitates the detection of the same astrophysical source over a broad spectrum of frequencies, from the early inspiral phase through to the final merger and ringdown phases. CE is designed as a 40 km L-shaped detector, and it is expected to be sensitive to GWs within the frequency range of 5 Hz to 5 kHz. In contrast, ET features a triangular configuration with 10 km long arms and is anticipated to operate within a frequency range of 1 Hz to 5 kHz. Here, we set the locations of CE and ET as Idaho (USA) and Cascina (Italy), respectively.

On the other hand, GWSat is a proposed Indian space-based interferometer designed to detect GWs at low frequencies, specifically within the frequency range of 0.1 to 5 Hz. We consider GWSat as having an L-shaped configuration with an arm length $L=100$km and an opening angle of 90 degrees. The configuration is considered such that the vertex spacecraft follows a geostationary orbit around the Earth (latitude 0 degrees). For GWSat, the wavelengths of the GWs, $\lambda_{\mathrm{GW}}$ corresponding to its sensitive frequency range (0.1 to 5Hz) are of the order of $10^5$ to $10^6$ km. These wavelengths are significantly larger than the detector's arm length $L$, which allows us to apply the long-wavelength approximation~\cite{Jaranowski:2005hz}. This long-wavelength approximation is also valid for ground-based detectors such as CE and ET. Therefore, we can use standard antenna pattern functions, similar to those utilized by ground-based detectors like LIGO, to characterize the sensitivity of GWSat, ET, and CE to gravitational waves coming from various directions.
To characterize the sensitivity of GWSat, we adopt a placeholder noise PSD, given by~\cite{PrivateCommMitra2024}:
\begin{equation}
    S_{n} (f) = (10^{-25} f^{-2} + 10^{-23}f^0 + 10^{-24} f^{-1})^2 \,\mathrm{Hz}^{-1}
\end{equation}

This PSD serves as a notional model and is not derived from detailed noise source calculations; however, it is intended to approximate the expected sensitivity profile for GWSat.
\subsection{Multiband visibility of IMRI system}\label{sec:Multiband_visibility}
To evaluate the detectability of an IMRI system across multiple GW detectors, we examine the SNR contribution of IMRI signals from different detectors.  Fig.~\ref{fig:noise_sensitivities} shows the strain sensitivities of the GWSat, CE, ET, and aLIGO detectors along with the characteristic strain of the GW signal from an IMRI system in the DM spike. For the CE detector, we use the PSD of the 40km baseline CE detector~\cite{Srivastava:2022slt}, while for ET, we utilize the ET-D sensitivity~\cite{Hild:2010id}. 
For aLIGO, we take the sensitivity curve of the advanced design sensitivity \texttt{aLIGOZeroDetHighPower}~\cite{aLIGO_design_sensitivity}.
The system under consideration consists of an IMRI with detector-frame companion masses of $400 \rm M_{\odot}$ and $1.4 \rm M_{\odot}$, located at a luminosity distance of $1 \text{Gpc}$. The detector-frame masses are the redshifted source-frame masses given by $m_z = m_{\rm source}(1+z)$. This figure illustrates the frequency evolution of the IMRI system over time as the binary components spiral inward. The system first becomes detectable in the GWSat band when the binary separation is large, accumulating a significant SNR before exiting the GWSat sensitivity range at 5 Hz. As it evolves, the system enters the ET and CE sensitivity ranges at 1 Hz and 5 Hz, respectively. Within the frequency interval between 1 Hz and 5 Hz, the GW signal becomes observable across both GWSat and ET detectors-facilitating simultaneous observation. The system transitions into the ET band approximately 2 hours before reaching ISCO and into the CE band around 1.7 minutes before $f_{\text{ISCO}}$, as indicated by the cyan solid line in Fig.~\ref{fig:noise_sensitivities}. Here we notice that the GW signal from the considered IMRI system will not be observable in the ground-based detector aLIGO.

\begin{figure}
    \centering
    \includegraphics[width=0.5\textwidth]{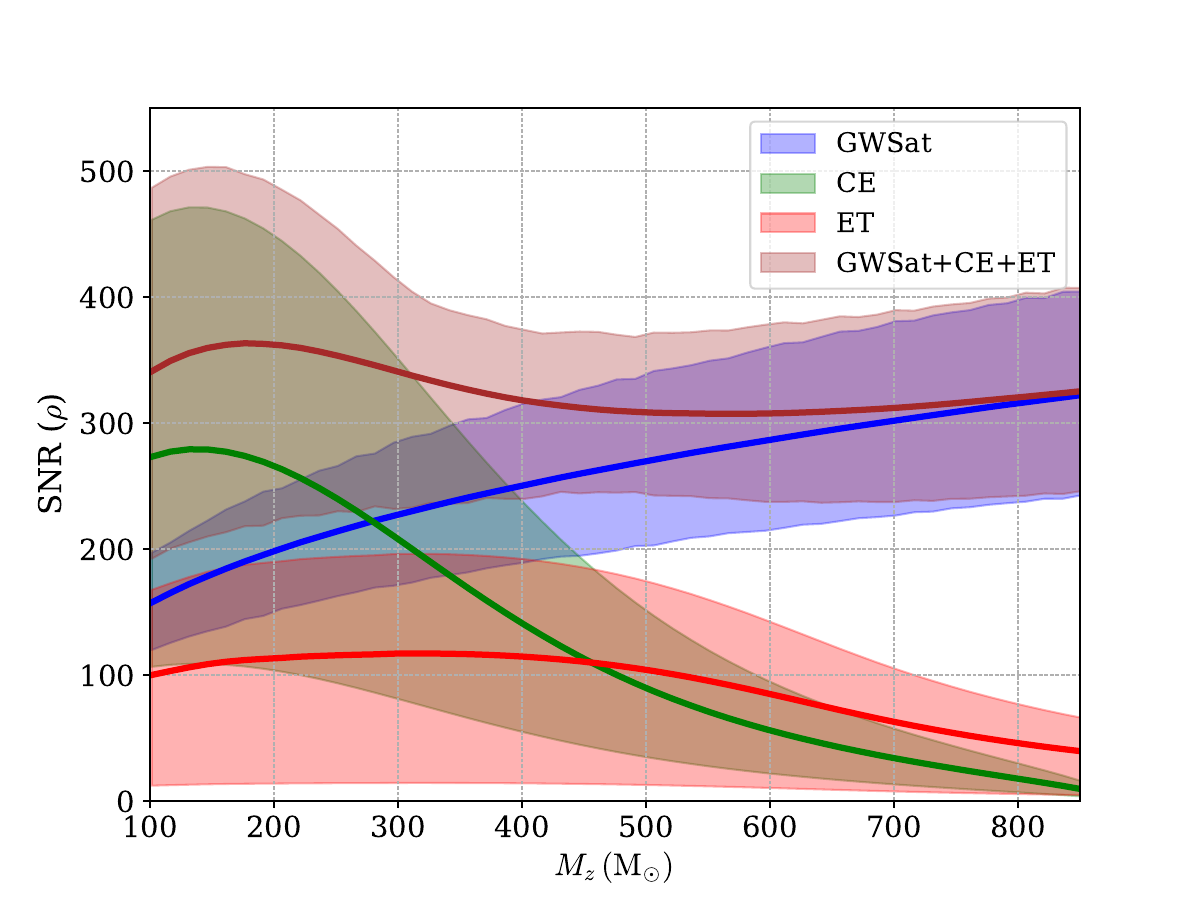}
    \caption{SNR from IMRIs in dynamic DM spike profile accumulated across the frequency bands of GWSat, CE, ET, as well as for combined CE+ET+GWSat observations up until $f_{\rm ISCO}$, as a function of detector-frame total mass $M_z$. The detector-frame secondary mass $m_2$ is set to 1.4$\mathrm{M}_{\odot}$. All the IMRI sources are considered to be located at 1 Gpc. The SNR is computed for 48 source locations uniformly distributed in the sky, assuming that the sources are oriented face-on with respect to the detector. The solid curves represent the average SNR across these source locations for each detector, while the shaded regions indicate the variation in SNR due to different sky positions.}
    \label{fig:SNR_for_different_M}
\end{figure}

\textcolor{black}{The SNR values for the IMRIs evolving in the DM environment, as a function of detector-frame total binary mass, $M_z$, are illustrated in Fig.~\ref{fig:SNR_for_different_M} for the GWSat, CE and ET detectors. Here, the mass of the primary component $m_1$ is varied while keeping $m_2$ fixed at 1.4$\mathrm{M}_{\odot}$. The SNR is computed for 48 uniformly distributed source locations across the sky. To simplify the analysis, we assume that all sources are face-on $(\iota = 0)$, as higher inclination angles $(\iota > 0)$ would introduce significant contributions from higher-order modes, which are not considered here. The polarization angle is arbitrarily set to $\pi/4$, and all sources are positioned at a distance of $D_L = 1$Gpc. The shaded band in the figure indicates the range of SNR values corresponding to different source locations for each detector.
The solid curves in the figure represent the average SNR across these source locations for various \( M_z \) values for each detector. Notably, in the CE and ET bands, the average SNR increases with $M_z$ until it reaches $150\mathrm{M}_{\odot}$ and $300\mathrm{M}_{\odot}$, respectively, as the signal strength rises with the total mass. After this point, the average SNR begins to decrease. This decrease is due to the inspiral phase of the IMRI system becoming progressively shorter in the CE and ET bands while becoming longer in the GWSat band as the masses surpass these values. Furthermore, for  $M_z$ values greater than $450 \, \mathrm{M}_{\odot}$, the average SNR in the ET band shows improvement compared to the CE band, showcasing the enhanced sensitivity of ET in the 1-5 Hz frequency range. Additionally, the SNR in the GWSat band begins to rise relative to the CE band beyond the $M_z = 300 \rm M_{\odot}$ as the visibility of the IMRI signal increases in the GWSat band. The average multiband SNR, which combines the contributions from all three detectors (CE+ET+GWSat), is denoted by the brown solid curve.}

The selection of the frequency bandwidth $(f_{\rm low}, f_{\rm high})$ for evaluating the integral in Eq.~(\ref{eq:Fisher_matrix}) is contingent on the particular detector or the GW event being considered. The frequency cut-offs ($f_{\mathrm{min}}$, $f_{\mathrm{max}}$) for GWSat is chosen to be in the range of (0.1, 5) Hz. For the terrestrial detectors CE and ET, the corresponding cut-off frequencies are set to (5, 100) Hz and (1, 100) Hz, respectively. The choice of the low-frequency cut-off for IMRIs in the GWSat band depends on the duration for which the signal is expected to persist within that frequency range. We adopt a one-year observation period in the space-based GWSat band, and the lower cutoff frequency for the integral in Eq.~(\ref{eq:Fisher_matrix}) is chosen as
\begin{equation}
    f_{\mathrm{low}} = \max{(f_{\mathrm{year}}, \: f_{\mathrm{min}})}, 
\end{equation}
where $f_{\mathrm{year}}$ is the GW frequency one year before ISCO. This approach ensures that the frequency does not fall below the detector’s sensitivity range by taking the maximum value between $f_{\rm year}$ and $f_{\rm min}$. The GW frequency at a specific observation time $T_{\mathrm{obs}}$ before ISCO is determined using the quadrupole formula for radiation damping, as given in Eq.~2.15 of \cite{Berti:2004bd}:
\begin{equation}
f_{\mathrm{year}} = 4.149 \times 10^{-5}\left[\frac{\mathcal{M}_z}{10^6 M_{\odot}}\right]^{-5 / 8}\left(\frac{T_{\mathrm{obs}}}{1 \mathrm{yr}}\right)^{-3 / 8}\,.
\end{equation}

The upper frequency limit for integration in Eq.~(\ref{eq:Fisher_matrix}) is chosen to be
\begin{equation}
    f_{\mathrm{high}} = \min{(f_{\mathrm{min}}, \: f_{\mathrm{ISCO}})}.
\end{equation}
where $f_{\mathrm{ISCO}}$ is the frequency corresponding to the innermost stable circular orbit for the Schwarzschild metric given by~\cite{Cutler:2007mi}
\begin{equation}
    f_{\mathrm{ISCO}} = \frac{1}{6^{3/2} \pi M_z } 
                        \simeq \frac{4397.16}{(M_z / \msun)}\: \Hz.
\end{equation}
%
\begin{figure*}[htbp!]
    \centering
    \includegraphics[width=0.95\textwidth]{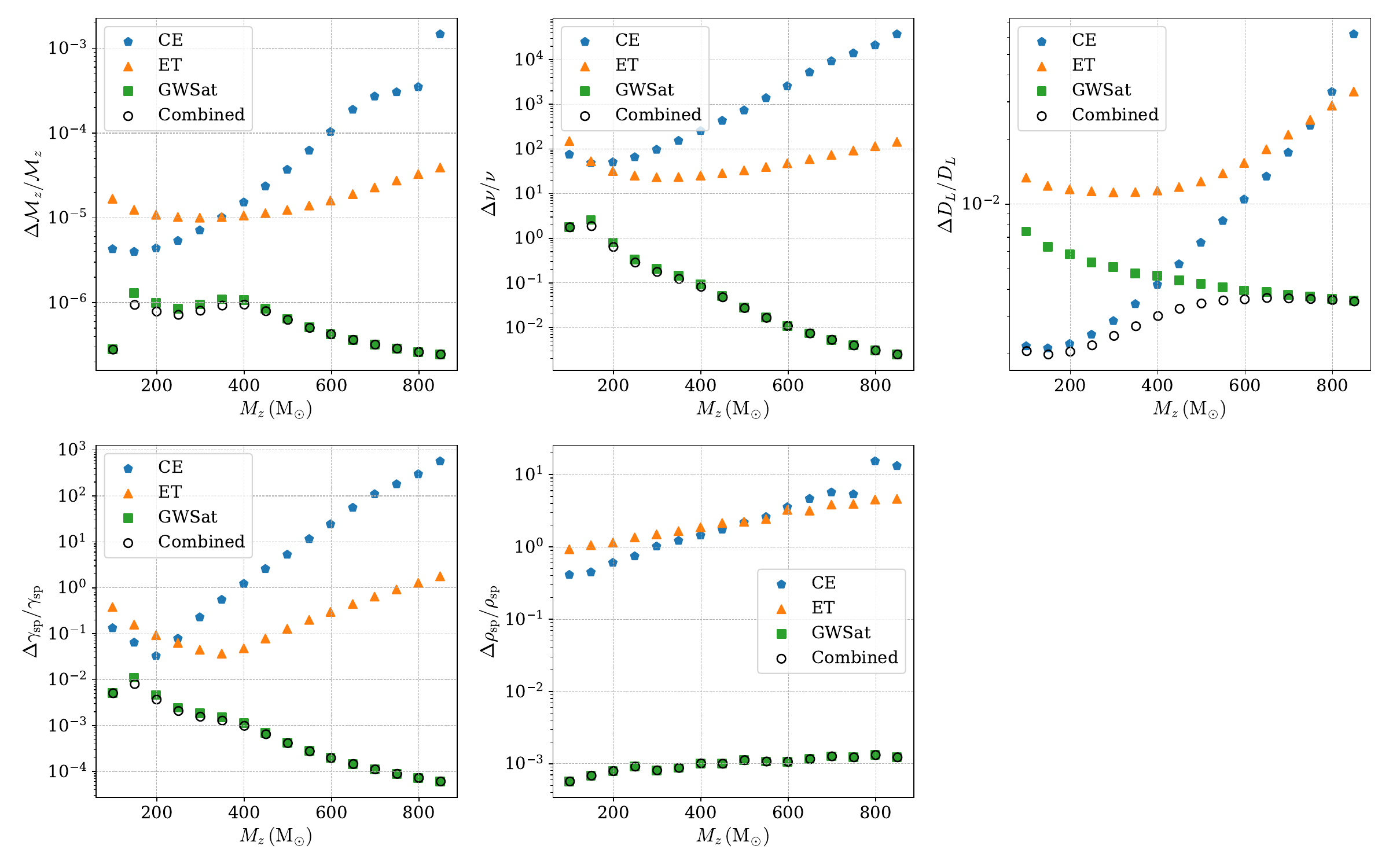}
    \caption{The error estimates for various binary parameters as well as DM spike parameters as a function of the total mass $M$ in the detector frame. The detector-frame secondary mass $m_2$ is kept constant at $1.4\mathrm{M}_{\odot}$, with all IMRIs systems considered to be situated at a luminosity distance of 1Gpc. The choices for the angles $(\alpha, \delta, \psi)$ are mentioned in \ref{sec:fisher_setup}. The fiducial values for the parameters $\rho_{\rm sp}$ and $\gamma_{\rm sp}$ are set at $226 \rm M_{\odot}/pc^3$ and $7/3$, respectively. The empty circle-shaped markers symbolize the combined bounds achieved through multiband observations incorporating GWSat, CE, and ET detectors. Notably, GWSat provides the most stringent constraints on  $\gamma_{\mathrm{sp}}$ and $\rho_{\mathrm{sp}}$, with uncertainties consistently remaining below $1\%$ across all values of $M_z$. Moreover, for $M_z > 400 \rm M_{\odot}$, the constraints on $\gamma_{\mathrm{sp}}$ improve further, tightening to below $0.1\%$. }
    \label{fig:one_sigma_bounds}
\end{figure*}
\begin{figure*}
    \centering
    \includegraphics[width=0.95\linewidth]{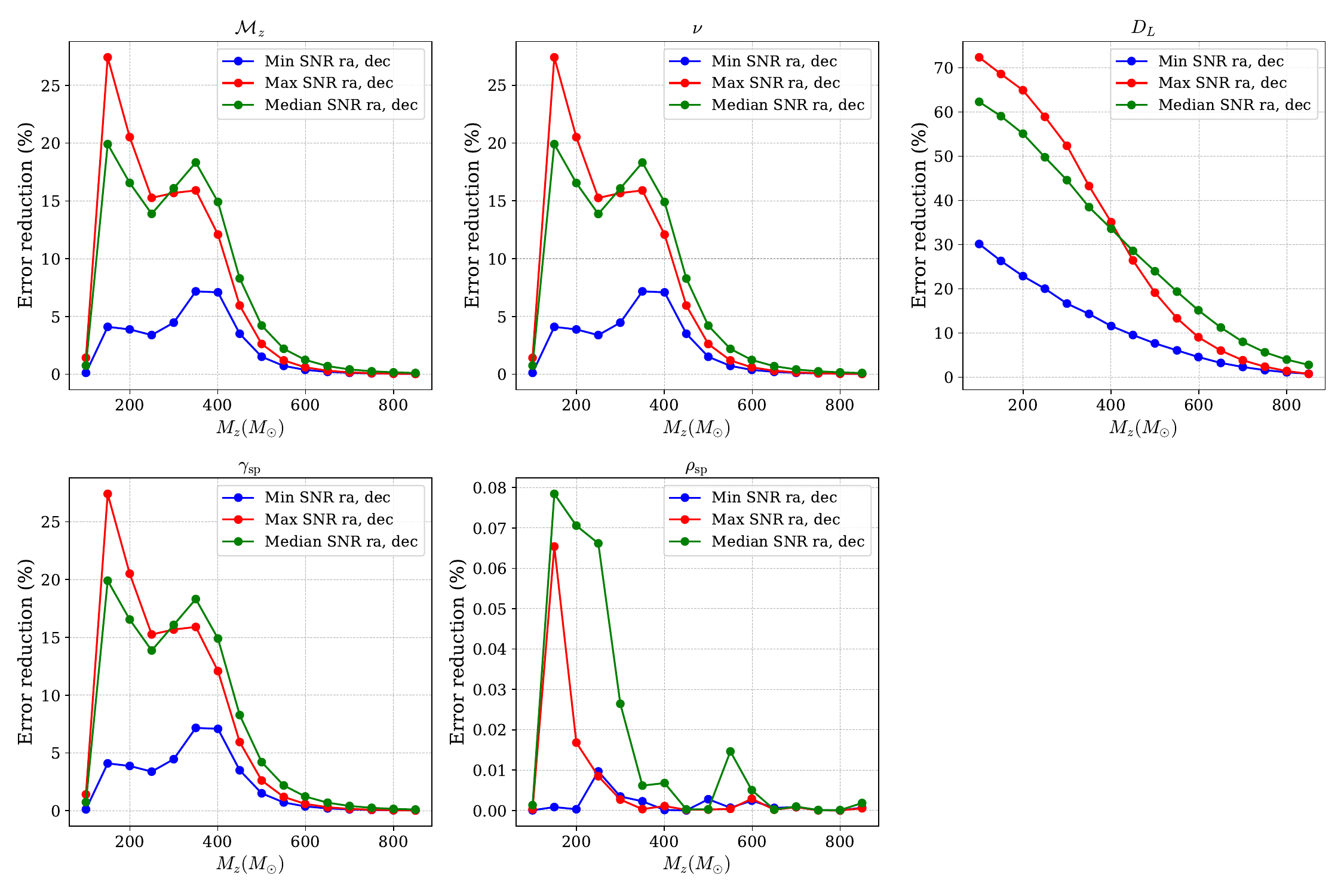}
    \caption{Percentage reduction in 1$\sigma$ bounds due to the inclusion of CE and ET to GWSat for the values of $(\alpha, \delta)$ for which we get maximum, minimum, and median SNR values in CE for a total mass of $300\rm M_{\odot}$. CE and ET enhance parameter constraints by over $10\%$ for $\mathcal{M}_z$, $\nu $, $D_L$, and $\gamma_{\rm sp}$ when  $M_z < 400 \mathrm{M}{\odot}$, while their impact on  $\rho_{\rm sp}$  remains negligible. For $M_z > 400 \mathrm{M}_{\odot}$, their contribution to error reduction decreases.}
    \label{fig:percentage_improvement}
\end{figure*}
\subsection{Setup for Fisher analysis}\label{sec:fisher_setup}
The parameter vector $\boldsymbol{\theta}$  for Fisher analysis depends on the waveform model used. To estimate the errors on parameters for an IMRI system in DM spike, we choose the following parameter space:
\begin{equation}
    \boldsymbol{\theta} = \{\mathcal{M}_z,\, \nu,\, D_L, \,\gamma_{\rm sp},\, \rho_{\mathrm{sp}}\}.
\end{equation}
We calculate statistical uncertainties for binary coalescence parameters as well as dark matter spike parameters for various IMRI systems using ground-based and space-based GW detectors. For this purpose, we employ \texttt{GWBENCH}~\cite{Borhanian:2020ypi}, a Python-based tool designed to compute the covariance matrix and the \(1\sigma\) errors on the parameters.  The plus and cross polarizations of the waveform for the dynamic DM spike scenario are incorporated within the \texttt{GWBENCH} framework for this analysis. The frequency cut-offs used for the integration in Eq.~(\ref{eq:Fisher_matrix}) for the estimation of errors are the same as mentioned in Section~\ref{sec:Multiband_visibility}. The fiducial value of the chirp mass $\mathcal{M}_z$ in the detector-frame is calculated using the detector-frame secondary mass $m_2=1.4\mathrm{M}_{\odot}$ and different values of detector frame primary BH mass $m_1$. \textcolor{black}{The fiducial value of $D_{L}$ and polarization angle are fixed to be 1Gpc and $\pi/4$, respectively. The fiducial values of the DM spike parameters are taken as $\rho_{\rm sp}=226 \rm M_{\odot}/pc^3$ and $\gamma_{\rm sp}=7/3$. For simplicity, the inclination angle is fixed at 0 degrees. As mentioned earlier, we divide the sky into 48 uniform regions and calculate the SNR for each pair of coordinates $(\alpha, \delta)$. To obtain the error estimates, we select the fiducial values for the right ascension and declination of the source that maximizes the SNR in CE for a total IMRI mass of $300\rm M_{\odot}$. Additionally, the \texttt{GWBENCH} facilitates the consideration of the effects of Earth’s rotation on detector antenna patterns while calculating the 1-$\sigma$ uncertainties on parameters.}
\section{Results} \label{sec:results}
\color{black} 
In this section, we present the results of our analysis in detail. Fig. \ref{fig:one_sigma_bounds} displays the parameter errors for an IMRI within a dynamic DM spike for individual detectors—GWSat, CE, and ET—as well as from multiband observations that combine GWSat, CE, and ET obtained using Fisher analysis. The constraints are shown for different values of the detector-frame total mass, $M_z$.

Our analysis demonstrates that the most stringent parameter constraints are obtained from the deciHz space-based detector GWSat since the IMRIs considered in this study spend the majority of their inspiral phase within the deciHz frequency band. In contrast, these systems complete fewer orbital cycles within the sensitivity bands of ground-based detectors like CE and ET, leading to comparatively weaker parameter constraints in those frequency ranges. The parameter estimation errors obtained from CE and ET increase for higher values of $M_z$ (i.e., ${M_z > 300 \rm  M_{\odot}}$ ) because of the reduced SNR of the signal for theses masses in CE and ET bands. 
GWSat provides the tightest constraints on the dark matter spike parameters, namely the spike index $\gamma_{\mathrm{sp}}$ and the density normalization $\rho_{\mathrm{sp}}$. The influence of DM on the binary evolution is most prominent in the deciHz band, and we find that the bounds on $\rho_{\mathrm{sp}}$ and $\gamma_{\mathrm{sp}}$ consistently remain below $1\%$ across all values of $M_z$, with the constraints on $\gamma_{\mathrm{sp}}$ tightening to below $0.1\%$ for $M_z > 400 \rm M_{\odot}$. The hollow black circles in Fig.~\ref{fig:one_sigma_bounds} illustrate the constraints derived from the combined analysis incorporating both space-based and terrestrial observations. 

Fig.~\ref{fig:percentage_improvement} illustrates the percentage reduction in the $1\sigma$ uncertainties $(\Delta \theta)$ for parameters $(\mathcal{M}_z,\, \nu, D_L, \, \gamma_{\mathrm{sp}},\, \rho_{\mathrm{sp}})$ when adding the contributions from the ground-based detectors (CE and ET) to the space-based GWSat detector. To assess the role of sky location in parameter estimation, we select three cases where CE records the maximum, minimum, and median SNR for a binary system with a total mass of  300 $\rm M_{\odot}$. 

At the sky location where CE achieves the maximum SNR, the inclusion of CE and ET leads to more than 15$\%$ reduction in the 1$\sigma$ uncertainty for the chirp mass $\mathcal{M}_z$, symmetric mass ratio $\nu$, luminosity distance $D_L$, and dark matter spike index $\gamma_{\rm sp}$ for systems with $M_z < 400 \rm M_{\odot}$. In contrast, at the minimum SNR location, the improvement is limited to ~5$\%$ for these systems. For higher total masses $( M_z > 400 \rm M_{\odot} )$, the contribution of CE and ET in reducing the error estimates of GWSat decreases. Notably, the inclusion of CE and ET has minimal impact on the constraints for $\rho_{\mathrm{sp}}$, with an improvement of less than $0.1\%$ regardless of the sky location.

It is observed that CE and ET improve the parameter bounds by more than $10\%$ for the chirp mass $\mathcal{M}_z$, symmetric mass ratio $\nu$, luminosity distance $D_L$, and dark matter spike index $\gamma_{\rm{sp}}$ for systems with $M_z < 400 \rm M_{\odot}$, compared to GWSat alone. It is important to note, however, that the improvement in $1\sigma$ bounds on $\rho_{\mathrm{sp}}$ from the inclusion of CE and ET remains minimal, at less than $0.01\%$. For higher total masses $(M_z > 400 \rm M_{\odot} )$, the contribution of CE and ET in reducing the error estimates of GWSat decreases due to lower SNR of these systems in their frequency band.
\color{black}

\section{Conclusion}\label{sec:conclusion}
\color{black}
We have investigated the capability of multiband gravitational wave observations to constrain the properties of dynamic dark matter spikes around IMRIs using the Fisher matrix analysis framework. Specifically, we studied IMRI systems embedded in dynamic dark matter spike profiles and examined how multiband observations, combining space-based and 3G ground-based detectors, can improve the estimation of both vacuum and dark matter parameters. We demonstrated that space-based detectors like GWSat play a crucial role in detecting the impact of dark matter, as IMRIs spend a significant fraction of their inspiral phase in the deciHz frequency band, where these effects accumulate over time.

The study focused on systems characterized by varying masses of the primary black hole and fixed secondary mass of $1.4\mathrm{M}_{\odot}$. Our findings show that GWSat alone provides the strongest constraints on dark matter parameters, particularly the spike density normalization $\rho_{\rm{sp}}$ and the power-law index $\gamma_{\rm{sp}}$, outperforming ground-based detectors due to its longer observation window at low frequencies. 

The addition of CE and ET improves parameter bounds for chirp mass $(\mathcal{M}_z)$, symmetric mass ratio $(\nu)$, luminosity distance $(D_L)$, and dark matter spike index $(\gamma_{\rm{sp}})$ by more than 15$\%$ for systems with $M_z < 400 \rm M_{\odot}$ at sky locations where CE records the highest SNR. However, this improvement is limited to approximately 5$\%$ at low-SNR locations, highlighting the dependence of parameter estimation on detector sensitivity across different regions of the sky. Moreover, the inclusion of CE and ET has a negligible impact on the constraints for $\rho_{\mathrm{sp}}$, with an improvement of less than 0.1$\%$, regardless of sky location. This result reinforces the necessity of space-based observations, as GWSat remains the dominant contributor to probing dark matter properties through gravitational waves.

However, there are certain caveats in this study. For simplicity, we assume that GWSat observes the IMRI system during the final year of its inspiral before the merger. This ensures a smooth transition of the system into the sensitive frequency band of ET, followed by CE, after it exits the GWSat band, thereby enabling multiband observations. This work primarily serves as a proof of principle, highlighting the potential of a deci-Hz detector in constraining dark matter properties using IMRI signals. 

We also assume that the signals detected by GWSat and ground-based observatories (CE and ET) can be confidently associated with the same IMRI system, allowing Fisher matrices to be combined for a multiband observation. In practice, identifying and matching signals across different detectors is a complex task, particularly for long-duration space-based observations where the data stream includes instrumental noise and foreground of overlapping signals. This issue has been discussed in previous studies (e.g., Ref~\cite{Lackeos:2023eub} in the context of LISA). 
While we do not address these practical challenges, recent studies such as sequential simulation-based inference methods like Peregrine \cite{Bhardwaj:2023xph}, have been proposed to resolve overlapping signals, while normalizing flow-based techniques~\cite{Langendorff:2022fzq} offer promising approaches for parameter estimation in complex detection scenarios. Adapting these methods to the deci-Hz band could significantly improve signal identification and enhance parameter constraints in multiband gravitational wave observations.
\color{black}

However, the IMRI systems are highly asymmetric, and the contribution of higher modes becomes important for more accurate estimations of the parameter uncertainty. We would like to explore the effects of the contribution of the higher modes to our analysis in a future study. Also, it will be interesting to extend our analysis for eccentric orbits and spinning binaries. Furthermore, we like to extend our studies to effective field theory models of dark matter (e.g., the one considered \cite{Bhattacharyya:2023kbh}). For this, we need to systematically compute the GW phase and carry out the Fisher analysis. Last but not least, we like to perform a thorough Bayesian analysis to get a more realistic bound on the dark matter parameter using multi-band observations. \textcolor{black}{Last but not least, it will be interesting to extend our studies for more general scenarios involving dynamic dark matter distribution e.g., \cite{Mukherjee:2023lzn, Kavanagh:2024lgq}}. We hope to report on some of these soon. 
\vspace{0.5cm}
%
\begin{acknowledgments}
A.B would like to thank the speakers and the participants of the (virtual) workshops “Testing Aspects of General Relativity-II”, “Testing Aspects of General Relativity-III”, and “New Insights into Particle Physics from Quantum Information and Gravitational Waves” at Lethbridge University, Canada funded by McDonald
Research Partnership-Building Workshop grant by McDonald Institute for valuable discussions. A.B. is supported by the SERB Core Research Grant (CRG/2023/005112) by DST-ANRF of India Govt.  A.B. also acknowledges the associateship program of the Indian Academy of Sciences, Bengaluru. D.T. would like to acknowledge the support of the Inspire fellowship from the Department of Science and Technology of India under fellowship number DST/INSPIRE Fellowship/[IF210643]. D.T. is grateful to Abhishek Sharma for valuable discussions and suggestions. D.T. also acknowledges Lalit Pathak for helpful discussions during the initial phase of this work. We sincerely thank the anonymous referee for their careful review and useful suggestions.
\end{acknowledgments}
\bibliography{reference}

\end{document}